\documentclass{article}

\usepackage[utf8]{inputenc}
\usepackage[T1]{fontenc}
\usepackage[letterpaper,margin=1in]{geometry}

\usepackage{assumptionsofphysics}
\usepackage{xcolor}
\usepackage{hyperref}
\usepackage{authblk}

\usepackage{biblatex}
\DeclareFieldFormat{doi}{DOI: \href{https://doi.org/#1}{#1}}
\def\N{\mathbb{N}}
\def\R{\mathbb{R}}
\def\C{\mathbb{C}}
\newcommand{\schwartz}[1][]{\mathcal{S}(\R^{#1},\C)}

\title{Quantum tomography on Schwartz space}
\author[1]{Tobias Thrien}
\author[2,3]{David Carfì}
\author[1]{Gabriele Carcassi}
\author[1]{Christine A. Aidala}
\affil[1]{University of Michigan, Ann Arbor, MI 48109, USA}
\affil[2]{University of Messina, 98166 Messina, Italy}
\affil[3]{University of California Riverside, Riverside, CA 92521, USA}
\date{\today}

\begin{document}

\maketitle

\begin{abstract}
    In this work we show that standard results in quantum tomography, which are typically proven in the context of Hilbert spaces, can analogously be established in the theory of Schwartz spaces. In particular, we will see that quantum states are equivalently characterized by their Wigner function, the characteristic of their Wigner function, and the probability distributions for a family of linear combinations of position and momentum. Physically, the latter can be understood as the evolution of its  probability distribution in position over time under, for example, free or harmonic evolution. The result is obtained with techniques that are more accessible to a physicist audience, and shows that, at least for tomography results, the full structure of Hilbert spaces is not necessary.
\end{abstract}

\section{Introduction}
In previous work~\cite{carcassi25}, some of us raised awareness of the difficulty of providing a physical justification for Hilbert spaces as a mathematical foundation for quantum mechanics. In particular, the completeness of the space with respect to the topology induced by the inner product necessarily introduces states with pathological physical properties. Efforts to reconstruct the theory of quantum mechanics based on physical requirements might therefore look for alternative targets. Here we follow up on our investigation of Schwartz spaces as a source of promising insights for alternative definitions and see whether we can establish analogue results completely within that theory.

Schwartz functions are smooth functions that decay rapidly at infinity, are dense in $L^2$ and have finite expectation values for all polynomials of position and momentum, which depend continuously on the state. These nice \emph{mathematical} properties considerably simplify their analysis by avoiding many of the domain issues common for (unbounded) operators on Hilbert spaces and other subtleties like functions only being defined almost everywhere. In this paper, we focus on the use of tomography as a powerful tool to analyze the \emph{physical} content encoded in Schwartz wavefunctions.

We will see that the state of a particle described by a Schwartz wavefunction can be uniquely identified (up to a global phase) by measuring the probability distribution of a family of linear combinations of position and momentum. This can also be interpreted as the evolution over time of the probability distribution in position, either freely or in the presence of a harmonic potential. While the latter can be achieved in finite time, the former would require infinite time. The probability densities in question are Schwartz functions as well, i.e., rapidly decaying at infinity. This result is analogous to the reconstruction of a probability distribution on classical phase space~\cite{cramer36,lyons18}.

While the general claim already exists (in different forms) in the literature (see e.g., Refs.~\cite{bertrand85,vogel89,kiukas08,cassinelli00,leonhardt96}), we provide here a short and straightforward proof, simplified by the properties of Schwartz spaces. In particular, we do not require any reference to the completeness of $L^2$. We hope that this makes the discussion more accessible to a broad physics community.

The paper starts by reviewing some mathematical prerequisites, including Wigner functions. The reconstruction then proceeds in the following steps.
\begin{figure}[ht]
    \centering
    $\psi \; \longleftrightarrow \; \mathcal{W}_\psi \longleftrightarrow \; \chi_\psi \; \longleftrightarrow \; \rho_\alpha \; \longleftrightarrow \; \rho_t$
    \caption{Sequence of reconstructions.}
    \label{fig:sequence}
\end{figure}
\begin{enumerate}
    \item The standard definition of a complex (wave-)function $\psi$ with certain regularity/decay properties.
    \item The well-known Wigner function $\mathcal{W}_\psi$, i.e., a quasiprobability distribution in phase space.
    \item The characteristic function $\chi_\psi$, i.e., expectation values of the unitary continuous operator $e^{i(aX+bP)}$ implementing translations in phase space.
    \item Probability distributions $\rho_\alpha$ for a family of linear combinations of the position and momentum operators $\cos \alpha \, X + \sin \alpha \, P$, which are all continuous operators.
    \item The probability distribution $\rho_t$ of the position operator evolving freely (or harmonically) in time.
\end{enumerate}

\section{Mathematical prerequisites}
In this section we provide the necessary prerequisites for what follows. First we introduce the central object of our analysis, namely Schwartz functions, which have a number of nice properties. Then we define Wigner functions as an alternative representation of quantum mechanical states and show the well-known result of how to recover the original wavefunction from its Wigner function, thereby showing that there exists a (constructive) one-to-one correspondence. Since the Wigner function is a special case of a more general transformation between operators and functions on phase space (see sec. \ref{sec:weyl-wigner}), it is more convenient to show how to reconstruct the kernel of the wavefunction, which is closer to the density matrix/operator of the state. 
Lastly we introduce a family of operators that will be used in the reconstruction later.

\subsection{Schwartz space}
\paragraph{Seminorms}
The Schwartz space $\schwartz[n]$ of smooth functions $f: \R^n \to \C$ and its topology are defined in terms of the following family of seminorms, indexed by multi-indices $\alpha, \beta \in \N^n$
\begin{equation}
    || f ||_{\alpha, \beta} = \sup_{x \in \R} \left| x^\alpha D^{\beta} f(x) \right|
\end{equation}
where $x^\alpha = x_1^{\alpha_1} x_2^{\alpha_2} \ldots x_n^{\alpha_n}$ are position operators and $D^{\beta} f = \partial_1^{\beta_1} \partial_2^{\beta_2} \ldots \partial_n^{\beta_n} f$ are (except for a factor of $-i \hbar$) momentum operators.

\paragraph{Topological vector space}
The Schwartz space is then
\begin{equation}
    \schwartz[n] = \left\{ f \in C^\infty(\R^n, \C) \, \middle| \, \forall a, b \in \N^n, || f ||_{\alpha, \beta} < \infty \right\}
\end{equation}
with the locally convex topology generated by all these semi-norms.

\paragraph{Inner product and Completeness}
We equip $\schwartz[n]$ with the $L^2$ inner product (and consequent norm)
\begin{equation}
    \langle \psi | \phi \rangle = \int_{\R^n} \psi^\ast\!(x) \, \phi(x) \, \mathrm{d}^n x
\end{equation}
which also induces another topology (which defines $L^2$), but is not used here. Note that $\schwartz[n]$ is complete w.r.t. to its own topology, but not w.r.t. to that of $L^2$. In fact, the unique completion of $\schwartz[n]$ in that topology is $L^2$. This makes $L^2$ a Hilbert space, while $\schwartz[n]$ is a pre-Hilbert space.

\paragraph{Physical seminorms}
Alternatively, one can use the following families of semi-norms closely related to the moments of position and momentum
\begin{align}
    || f ||^{L^1}_{\alpha, \beta} &= \int_{\R^n} \left| x^\alpha D^{\beta} f(x) \right| \, \mathrm{d}^n x \\
    || f ||^{L^2}_{\alpha, \beta} &= \left( \int_{\R^n} \left| x^\alpha D^{\beta} f(x) \right|^2 \, \mathrm{d}^n x \right)^{1/2}
\end{align}
which will analogously define the same set and topology~\cite{brief010}.

\paragraph{Fourier transform}
The Schwartz space is closed under the Fourier transform $\mathcal{F}: \schwartz[n] \to \schwartz[n]$
\begin{equation}
    (\mathcal{F} \psi) (p) = \frac{1}{(2 \pi \hbar)^n} \int_{\R^n} e^{-\frac{i}{\hbar} p x} \, \psi(x) \, \mathrm dx
\end{equation}
and its inverse
\begin{equation}
    (\mathcal{F}^{-1} \psi) (x) = \int_{\R^n} e^{\frac{i}{\hbar} p x} \, \psi(p) \, \mathrm dp \; .
\end{equation}
The same holds for partial Fourier transforms like
\begin{equation}
    (\mathcal{F}_1 \psi) (p_1, x_2) = \frac{1}{2 \pi \hbar} \int_{\R} e^{-\frac{i}{\hbar} p_1 x_1} \, \psi(x_1, x_2) \, \mathrm dx_1 \; .
\end{equation}

\paragraph{Operator algebra}
All operators in this paper are everywhere defined on $\schwartz[n]$, linear and continuous. These form the algebra $\mathcal{L}(\mathcal{S})$ w.r.t. composition. In this context symmetric and self-adjoint are equivalent, namely for all $\psi, \phi \in \schwartz[n]$
\begin{equation}
    \langle O \psi | \phi \rangle = \langle \psi | O \phi \rangle \; .
\end{equation}

\subsection{Wigner functions}
\begin{defn}[Wigner function]\label{def:wigner}
    For a function $\psi \in \schwartz$, its Wigner function $\mathcal{W}_\psi : \R^2 \to \R$ is given by
    \begin{equation}\label{eq:wigner}
        \mathcal{W}_\psi(x, p) = \frac{1}{2\pi \hbar} \int_{\R} e^{-\frac{i}{\hbar} p y} \, \psi\left(x + \frac{y}{2} \right) \, \overline{\psi\left( x - \frac{y}{2} \right)} \, \mathrm dy \, .
    \end{equation}
    The Wigner function is a real Schwartz function in $\schwartz[2]$ (see Prop. 1.92 in \cite{folland16}).
\end{defn}

\begin{defn}[Kernel]\label{def:kernel}
    The kernel $K_\psi : \R^2 \to \C$ of a function $\psi \in \schwartz$ is given by
    \begin{equation}\label{eq:kernel}
        K_\psi(x, y) = \psi(x) \, \overline{\psi(y)} \, .
    \end{equation}
\end{defn}

\begin{prop}
    The kernel $K_\psi : \R^2 \to \C$ of a Schwartz function $\psi \in \schwartz$ is again a Schwartz function, i.e., $K_\psi \in \schwartz[2]$.
\end{prop}
\begin{proof}
    Since $\psi \in C^\infty(\R, \C)$ also $\overline{\psi} \in C^\infty(\R, \C)$ and their product. The norm factors
    \begin{align}
        || K_\psi ||_{\alpha, \beta} &= \sup_{x \in \R^2} \left| x^\alpha \, D^{\beta} \psi(x_1) \, \overline{\psi(x_2)} \right| \\
        &= \sup_{x \in \R^2} \left| x_1^{\alpha_1} \, D^{\beta_1} \psi(x_1) \, x_2^{\alpha_2} \, D^{\beta_2} \overline{\psi(x_2)} \right| \\
        &= \sup_{x \in \R^2} \left| x_1^{\alpha_1} \, D^{\beta_1} \psi(x_1) \, \middle| \middle| x_2^{\alpha_2} \, D^{\beta_2} \overline{\psi(x_2)} \right| \\
        &\leq \left( \sup_{x_1 \in \R} \left| x_1^{\alpha_1} \, D^{\beta_1} \psi(x_1) \right| \right) \left( \sup_{x_2 \in \R} \left| x_2^{\alpha_2} \, D^{\beta_2} \overline{\psi(x_2)} \right| \right) \\
        &\leq || \psi ||_{\alpha_1, \beta_1} \, || \overline{\psi} ||_{\alpha_2, \beta_2} < \infty
    \end{align}
\end{proof}

\begin{thrm}[Kernel reconstruction]\label{thrm:kernel}
    Given the Wigner function $\mathcal{W}_\psi : \R^2 \to \R$ of some function $\psi \in \schwartz$ we can reconstruct its kernel $K_\psi$ by
    \begin{equation}\label{eq:kernel_reconstruction}
        K_\psi(x,y) = \int_{\R} e^{\frac{i}{\hbar} p (x - y)} \, \mathcal{W}_\psi\left( \frac{x + y}{2}, p \right) \, \mathrm dp \, .
    \end{equation}
\end{thrm}
\begin{proof}
    Plugging def.~\ref{def:wigner} into eq.~\eqref{eq:kernel_reconstruction} yields\footnote{Note that the double integral is not absolutely convergent since the modulus of the integrand does not depend on $p$. Fubini's theorem therefore does not allow us to change the order of integration, and the Lebesgue integral over $\R^2$ does not exist.}
    \begin{align}
        K_\psi(x,y) &= \frac{1}{2\pi \hbar} \int_{\R} e^{\frac{i}{\hbar} p (x - y)} \left( \int_{\R} e^{-\frac{i}{\hbar} pz} \, \psi\left(\frac{x + y + z}{2} \right) \, \overline{\psi\left( \frac{x + y - z}{2} \right)} \, \mathrm dz \right) \mathrm dp
    \end{align}
    where the inner integral over $z$ is the Fourier transform of $a(z) = \psi\left(\frac{x + y + z}{2} \right) \, \overline{\psi\left( \frac{x + y - z}{2} \right)}$ leaving
    \begin{equation}
        K_\psi(x,y) = \int_{\R} e^{\frac{i}{\hbar} p (x - y)} \, \hat{a}(p) \, \mathrm dp = a(x - y)
    \end{equation}
    which then simplifies, after inserting the definition of $a$ again, to
    \begin{equation}
        K_\psi(x,y) = a(x - y) = \psi\left(\frac{x + y + x - y}{2} \right) \, \overline{\psi\left( \frac{x + y - (x - y)}{2} \right)} = \psi(x) \, \overline{\psi(y)}
    \end{equation}
    as expected.
\end{proof}

Alternatively we can prove this using a generalized notion of orthogonality for the de Broglie family $e^{\frac{i}{\hbar} px}$. See appendix \ref{app:deBroglie} for more.

\begin{thrm}[Wavefunction reconstruction]
    Given the kernel $K_\psi : \R^2 \to \C$ of some (normalized) function $\psi \in \schwartz$, we can find a function $\varphi \in \schwartz$ with the same kernel. For that, find a point $y_0$ where $K_\psi$ does not vanish and fix the phase convention
    \begin{equation}
        \varphi(y_0) = \sqrt{K_\psi(y_0,y _0)} > 0 \, .
    \end{equation}
    Then set
    \begin{equation}
        \varphi(x) = \frac{K_\psi(x, y_0)}{\sqrt{K_\psi(y_0, y_0)}} \, .
    \end{equation}
    The (re)constructed function $\varphi$ only differs from $\psi$ by a phase, i.e., $\varphi = e^{i \theta} \psi$.
\end{thrm}
\begin{proof}
    Notice that diagonal entries in the kernel are necessarily real and the square root is therefore well-defined, real and positive since we picked a point $y_0$ where the kernel does not vanish, which must exist for the kernel of anything but the zero function.
    \begin{align}
        \varphi(x) = \frac{K_\psi(x, y_0)}{\sqrt{K_\psi(y_0, y_0)}} = \frac{\psi(x) \, \overline{\psi(y_0)}}{\sqrt{\psi(y_0) \, \overline{\psi(y_0)}}} = \frac{\overline{\psi(y_0)}}{|\psi(y_0)|} \psi(x) = e^{i \theta} \psi(x)
    \end{align}
\end{proof}

\begin{thrm}[Reconstruction]
    From the previous two theorems it immediately follows that we can reconstruct any (normalized) function $\psi \in \schwartz$ from its Wigner function $\mathcal{W}_\psi$ up to a global phase $e^{i \theta}$.
\end{thrm}

\subsection{Weyl--Wigner Correspondence}\label{sec:weyl-wigner}
\begin{defn}
    For a function $w: \R^2 \to \C \in \mathcal{O}_M(\R^2)$ its Weyl quantization $\mathcal{O}_w: \schwartz \to \schwartz$ is given by
    \begin{equation}
        (\mathcal{O}_w \varphi)(x) = \frac{1}{2\pi \hbar} \int_{\R} e^{\frac{i}{\hbar} px} \left( \int_{\R} e^{-\frac{i}{\hbar} py} \, w\left( \frac{x + y}{2}, p \right) \, \varphi(y) \, \mathrm dy \right) \mathrm dp \, .
    \end{equation}
    Compare eq. (2.3) in \cite{folland16}.
\end{defn}

\begin{thrm}[Weyl--Wigner identity]\label{thrm:weyl-wigner}
    The expectation values are preserved when transforming operators to functions on phase space and wave functions to Wigner functions, i.e.,
    \begin{equation}
        \langle \psi | \mathcal{O}_w \psi \rangle = \iint_{\R^2} w(x, p) \, \mathcal{W}_\psi(x, p) \, \mathrm dx \, \mathrm dp \, .
    \end{equation}
    See Prop. 2.5 in \cite{folland16}.
\end{thrm}

\subsection{Weyl-Heisenberg Operators}
We consider the family of unitary operators $W(a, b): \schwartz \to \schwartz$ defined for any $a, b \in \R$ as
\begin{equation}
    W(a, b) = \exp \left( \frac{i}{\hbar} (a X + b P) \right)
\end{equation}
where the position operator $X: \schwartz \to \schwartz$ and the momentum/derivative operator $P: \schwartz \to \schwartz$ are defined respectively as
\begin{align}
    (X \psi) (x) &= x \, \psi(x) \\
    (P \psi) (x) &= -i \hbar \frac{\mathrm{d} \psi}{\mathrm{d}x}(x) \quad .
\end{align}
Both of these operators are defined everywhere, continuous in the Schwartz topology, and symmetric, which is the same as self-adjoint in this case on $\schwartz$. The exponential function can be defined in terms of its power series, which would not be possible on $L^2$.

Alternatively, they can be defined via translations and modulations.
\paragraph{Translation}
For $b \in \R$, define $T_b : \schwartz \to \schwartz$ as
\begin{equation}
    (T_b \varphi)(x) = \varphi(x + b) \, .
\end{equation}

\paragraph{Modulation}
For $a \in \R$, define $M_a : \schwartz \to \schwartz$ by
\begin{equation}
    (M_a \varphi)(x) = e^{\frac{i}{\hbar} a x} \varphi(x) \, .
\end{equation}

\begin{defn}
    For $(a,b) \in \R^2$, define the Weyl--Heisenberg operator on $\schwartz$ by
    \begin{equation}
        W(a, b) = e^{\frac{i}{2 \hbar} a b} M_a T_b \, .
    \end{equation}
    Equivalently, for $\varphi \in \schwartz$,
    \begin{equation}
        (W(a, b) \varphi)(x) = e^{\frac{i}{\hbar} a (x + \frac{b}{2})} \, \varphi(x + b) \, .
    \end{equation}
\end{defn}

\section{Reconstructions}
We now proceed with the next steps in our reconstruction.

\subsection{Characteristic function}
\begin{defn}
    The characteristic function $\chi_\psi: \R^2 \to \C$ of a function $\psi \in \schwartz$ is
    \begin{equation}
        \chi_\psi(a, b) = \langle \psi | W(a, b) \psi \rangle \, .
    \end{equation}
\end{defn}

\begin{prop}[Fourier relation]\label{prop:fourier}
    For $\psi \in \schwartz$,
    \begin{equation}
        \chi_\psi(a, b) = \iint_{\R^2} e^{\frac{i}{\hbar}(ax + bp)} \, \mathcal{W}_\psi(x, p) \, \mathrm dx \, \mathrm dp \, ,
    \end{equation}
    and conversely,
    \begin{equation}
        \mathcal{W}_\psi(x, p) = \frac{1}{(2\pi \hbar)^2} \iint_{\R^2} e^{-\frac{i}{\hbar} (ax + bp)} \, \chi_\psi(a, b) \, \mathrm da \, \mathrm db \, .
    \end{equation}
\end{prop}
\begin{proof}
    The first relationship follows directly from Theorem~\ref{thrm:weyl-wigner} since $W(a, b)$ is the Weyl quantization for fixed $a$ and $b$ of
    \begin{equation}
        w(x, p) = e^{\frac{i}{\hbar}(ax + bp)}
    \end{equation}
    To see this explicitly:
    \begin{align}
            (W(a, b) \, \varphi)(x)
            &= \frac{1}{2\pi \hbar} \int_{\R} e^{\frac{i}{\hbar} px} \left( \int_{\R} e^{-\frac{i}{\hbar} py} \, e^{\frac{i}{\hbar}(a \frac{x + y}{2} + bp)} \, \varphi(y) \, \mathrm dy \right) \mathrm dp \\
            &= e^{\frac{i}{\hbar} \frac{a}{2}x} \int_{\R} e^{\frac{i}{\hbar} p(x + b)} \left( \frac{1}{2\pi \hbar} \int_{\R} e^{-\frac{i}{\hbar} (p - \frac{a}{2})y} \, \varphi(y) \, \mathrm dy \right) \mathrm dp \\
            &= e^{\frac{i}{\hbar} \frac{a}{2}x} \int_{\R} e^{\frac{i}{\hbar} p(x + b)} \Tilde{\varphi}\left(p - \frac{a}{2}\right) \mathrm dp \\
            &= e^{\frac{i}{\hbar} \frac{a}{2}x} \int_{\R} e^{\frac{i}{\hbar} (p' + \frac{a}{2})(x + b)} \Tilde{\varphi}\left(p'\right) \mathrm dp' \\
            &= e^{\frac{i}{\hbar} (\frac{a}{2}x + \frac{a}{2}(x+b))} \int_{\R} e^{\frac{i}{\hbar} p'(x + b)} \Tilde{\varphi}\left(p'\right) \mathrm dp' \\
            &= e^{\frac{i}{\hbar} a (x + \frac{b}{2})} \, \varphi(x + b)
    \end{align}
    The inverse relationship follows immediately, once we recognize the first one to be the Fourier transform.
\end{proof}

\begin{coro}
    Since the Fourier transform preserves the Schwartz space we conclude that the characteristic function $\chi_\psi$ of a Schwartz function $\psi$ is again a Schwartz function.
\end{coro}

\begin{thrm}\label{thrm:reconstruction}
    The expectation values of $W(a, b)$
    \begin{equation}
        \langle \psi | W(a, b) \psi \rangle \,
    \end{equation}
    uniquely define the state $\psi \in \schwartz$ up to a global phase.
\end{thrm}
\begin{proof}
    The expectation values, i.e., values of the characteristic function, are the Fourier transform of the Wigner function. Since the Fourier transform is invertible and the Wigner function uniquely determines the state, we can also recover the state from these expectation values.
\end{proof}

\subsection{Probability distributions}
We will define the probability distribution of linear combinations of $X$ and $P$ like\footnote{For simplicity we ignore the units here and consider $X$ and $P$ to be dimensionless (or measured in a particular fixed unit).}
\begin{equation}
    \mathcal{X}(\alpha) = \cos \alpha \, X + \sin \alpha \, P \, .
\end{equation}

\begin{prop}
    This parametrized curve can be generated by the operator $A: \schwartz \to \schwartz$, defined as\footnote{The generator coincides with the Hamiltonian of the harmonic oscillator.}
    \begin{align}
        A = \frac{1}{2} \left( X^2 + P^2 \right) \, ,\quad
        U_\alpha = e^{-\frac{i}{\hbar} \alpha A} \\
        \mathcal{X}(\alpha) = U_\alpha X U_\alpha^\dagger = \cos \alpha \, X + \sin \alpha \, P \, . \label{eq:solution}
    \end{align}
    Note that there are no domain problems for $A$ since $X$ and $P$ are already defined everywhere and map back to $\schwartz$.
\end{prop}
\begin{proof}
    $X(\alpha)$ satisfies the following 2nd order ODE:
    \begin{equation}
    		\frac{\mathrm d^2X}{\mathrm d\alpha^2} (\alpha) = \left[ \left[ X(\alpha) ,\frac{i}{\hbar}  A \right] , \frac{i}{\hbar} A \right] = \left[ P(\alpha) , \frac{i}{\hbar} A \right] = -X(\alpha)
    \end{equation}
    which is solved by the Ansatz: $X(\alpha) = a(\alpha) X + b(\alpha) P$. Then $a$ and $b$ satisfy
	\begin{align}
		\frac{\mathrm d^2a}{\mathrm d\alpha^2} &= -a(\alpha) &
		\frac{\mathrm d^2b}{\mathrm d\alpha^2} &= -b(\alpha) \\
		\left. \frac{\mathrm da}{\mathrm d\alpha} \right|_{\alpha=0} &= 0 &
		\left. \frac{\mathrm db}{\mathrm d\alpha} \right|_{\alpha=0} &= 1 \\
		a(0) &= 1 &
		b(0) &= 0
	\end{align}
	and have well-known solutions:
	\begin{align}
		a(\alpha) = \cos(\alpha) \quad b(\alpha) = \sin(\alpha) \\
		\implies \mathcal{X}(\alpha) = \cos(\alpha) X + \sin(\alpha) P
	\end{align}
	See appendix \ref{app:harmonicoscillatorheisenberg} for more detailed calculations.
\end{proof}

Similarly, we can transform the initial probability distribution in $X = \mathcal{X}(0)$ for a given state $\psi \in \schwartz$ as
\begin{equation}
    \rho_0(q) = | \psi(q) |^2 \quad \to \quad \rho_\alpha(q) = | (U_\alpha \psi)(q) |^2
\end{equation}

\begin{thrm}\label{thrm:radon}
    The Fourier transform $\hat{\rho}_\alpha$ of the probability distribution $\rho_\alpha$ of $\mathcal{X}(\alpha)$ (as defined above) is related to the characteristic function $\chi_\psi$ as follows
    \begin{equation}
        \hat{\rho}_\alpha(s)
        := \int_\R e^{\frac{i}{\hbar} s q} \, \rho_\alpha(q) \, \mathrm dq
        = \chi_\psi(s \cos \alpha, s \sin \alpha) \, .
    \end{equation}
\end{thrm}
\begin{proof}
    Let us call $\psi_\alpha = U_\alpha \psi$.

    \begin{align}
            \hat{\rho}_\alpha(s)
            &= \int_\R \bar{\psi}_\alpha(q) e^{\frac{i}{\hbar} s q} \, \psi_\alpha(q) \, \mathrm dq \\
            &= \langle U_\alpha \psi | e^{\frac{i}{\hbar} s X} U_\alpha \psi \rangle \\
            &= \langle \psi | U_\alpha^\dagger e^{\frac{i}{\hbar} s X} U_\alpha \psi \rangle \\
            &= \langle \psi | e^{\frac{i}{\hbar} s U_\alpha^\dagger X U_\alpha} \psi \rangle \\
            &= \langle \psi | e^{\frac{i}{\hbar} s \mathcal{X}(\alpha)} \psi \rangle \\
            &= \langle \psi | e^{\frac{i}{\hbar} s (\cos \alpha \, X + \sin \alpha \, P)} \psi \rangle \\
            &= \langle \psi | W(s \cos \alpha, s \sin \alpha) \psi \rangle \\
            &= \chi_\psi(s \cos \alpha, s \sin \alpha)
    \end{align}
\end{proof}

Note that $\hat{\rho}_\alpha$ is again a Schwartz function for every $\alpha$.

\begin{thrm}
    Given the probability distributions $\rho_\alpha$ with $\alpha \in [0, \pi)$ of the observables $\mathcal{X}(\alpha) = \cos \alpha X + \sin \alpha P$ for a state $\psi \in \schwartz$, as defined above, we can reconstruct the original state.
\end{thrm}
\begin{proof}
    As shown in theorem \ref{thrm:radon}, we can reconstruct the characteristic function $\chi_\psi$ from $\hat{\rho}_\alpha$, which we obtain from $\rho_\alpha$ using the inverse Fourier transform.

    Then theorem \ref{thrm:reconstruction} tells us that we can reconstruct the original state from $\chi_\psi$.
\end{proof}

\subsection{Physical interpretation}
As has been noted above the generator of the family is the Hamiltonian of the harmonic oscillator. Physically, this means that the family can be interpreted as the position operator evolving under the harmonic potential in the Heisenberg picture, and the parameter $\alpha$ represents time. However, this is not the only Hamiltonian one can use for tomography.

For free evolution, we know classically $x(t) = x(0) + \frac{p}{m}t = x(0) + \frac{t}{m} p$. Setting $\tan(\alpha) = \frac{t}{m}$ we can see that this curve of operators corresponds to observing the position $X$ over time $t$ during \emph{free} evolution.
\begin{equation}
    \Tilde{\mathcal{X}}(t) = \frac{1}{\cos(\alpha(t))} \mathcal{X}(\alpha(t)) = X + \tan(\alpha) P = X + \frac{t}{m} P
\end{equation}
Note that while the point $\alpha = \pi / 2$ is never reached we can get arbitrarily close and since expectation values are continuous the single point does not contain any further data.

To see this explicitly, define the Hamiltonian for free evolution.
\begin{equation}
    H = \frac{1}{2m} P^2
\end{equation}
It generates time evolution via
\begin{equation}
    U_t = e^{-\frac{i}{\hbar} H t}
\end{equation}
and therefore
\begin{equation}
    \psi_t = U_t \psi
\end{equation}

The probability density for the distribution of position $X$ in the state $\phi_t$ at time $t$ is then
\begin{equation}
    \rho_t(x) = | \psi_t (x) |^2 = | (U_t \psi)(x) |^2 = | (e^{-\frac{i}{\hbar} H t} \psi)(x) |^2 = | (e^{-\frac{i}{2 m \hbar} P^2 t} \psi)(x) |^2 \; .
\end{equation}
In the Heisenberg picture this corresponds to the distribution of the observable $U_t X U_t^\dagger$. See appendix \ref{app:freeheisenberg} for a derivation.

\section{Conclusion}
We presented a short proof of several tomographic characterizations of Schwartz wavefunctions. The proof requires only basic tools from analysis, which we hope will make it accessible to a broad audience. The characterization in terms of the probability distribution $\rho_\alpha$ at specific angles $\alpha$ in phase space has a precise analogue in classical mechanics. Under harmonic evolution, probability densities on phase space rotate around the origin (for some choice of units). The marginals for all rotation angles $\alpha$ uniquely determine the full phase-space measure by application of the Cramér–Wold theorem~\cite{cramer36,lyons18}.

This result sits within our larger goal of understanding what vector space representation of quantum states can be both mathematically precise and physically meaningful. The result shows that, at least for quantum tomography, the Schwartz spaces provide analogous results that are independent of the Hilbert space theory. While Schwartz spaces are limited to quantum systems describing only positional degrees of freedom of particles, and therefore cannot be the ultimate solution, studying Schwartz analogues may pave the way to find better alternatives to the standard Hilbert space formalism.

\printbibliography
\pagebreak

\appendix
\section{Appendix}
\subsection{Further mathematical prerequisites}
We provide some further properties of Schwartz spaces that allow us to use different techniques that are much closer to common practice in physics to prove some of the results in this paper.

\paragraph{Multipliers}
Schwartz spaces are closed under multiplication by any function in $\mathcal{O}_M(\R^n)$, the space of slowly increasing smooth function, i.e., all $f \in C^\infty(\R^n, \C)$, s.t.
\begin{equation}
    \forall \alpha \in \N^n, \exists k \in \N, C \ge 0, \forall x \in \R^n : \left| \partial^\alpha f(x) \right| \le C \left(1 + |x|^2 \right)^k \; ,
\end{equation}
which are \emph{precisely} those functions that preserve Schwartz spaces under multiplication.

\paragraph{Tempered distributions}
In contrast to $L^2$, which is self-dual, the dual\footnote{By dual space we mean the topological dual, i.e., all \textit{continuous} linear functionals from $\schwartz[n]$ to $\R$.} space of $\mathcal{S}$ is much bigger and contains elements like the Dirac distribution\footnote{Since $\delta$ is a distribution and not a function, the integral in eq.~\eqref{eq:dirac} is merely convenient notation and should not be taken literally without further qualifications.}
\begin{equation}\label{eq:dirac}
    \delta_x[f] = f(x) = \int_{\R^n} \delta(y - x) f(y) \, \mathrm{d}^n y
\end{equation}
which cannot be defined on $L^2$ since it does not contain functions but equivalence classes of functions. The elements of this dual space $\mathcal{S}'$ are called \textit{tempered distributions}.

\paragraph{Integral and Dirac measure}
Taking the integral above literally requires some further qualifications, since there is no function $\delta : \R^n \to \R$ that would satisfy eq.~\eqref{eq:dirac}. However, $\delta_x \in \mathcal{S}'$ can be multiplied by $f \in \schwartz[n]$ (or in general $\mathcal{O}_M(\R^n)$) to give another tempered distribution, which can be evaluated at the constant function $1$.
\begin{equation}
    \delta_x[f] = \delta_x[f \cdot 1] = (\delta_x \cdot f)[1] = \int_{\R^n} 1 \, (f \delta_x)
\end{equation}
The first multiplication $\cdot$ is between an element of $\mathcal{O}_M(\R^n)$ and $\schwartz[n]$, while the second multiplication is between $\mathcal{S}'$ and $\schwartz[n]$ and defined in terms of the first, just as written. Note that even though $1 \notin \schwartz[n]$ we can evaluate $\delta_x \cdot f$ on any $\mathcal{O}_M(\R^n)$ function and any indicator function $1_U$ of some Borel set $U \subseteq \R^n$. This defines a measure
\begin{equation}
    \mu(U) := (\delta_x \cdot f) [ 1_U ]
\end{equation}
which can be integrated over $\R^n$ to give the same result.
\begin{equation}
    \delta_x [ f ] = \int_{\R^n} \mathrm d\mu
\end{equation}

\paragraph{Approximation by integrals}
Another justification for using the integral notation is that $\mathcal{S}$ is dense in $\mathcal{S}'$, i.e., every linear functional in $\mathcal{S}'$ can be approximated by integrals of the form
\begin{equation}
    \int_{\R^n} f_i(x) \, g(x) \, \mathrm dx \; \to \; F[g],
\end{equation}
where $f_i \in \mathcal{S}$. The $\delta_x$ distribution for example can be approximated by Gaussian probability density functions with mean $x$ and standard deviation $\sigma \to 0$.

\subsection{Alternative proof using orthogonal de Broglie family}\label{app:deBroglie}
This is an alternative proof of theorem \ref{thrm:kernel} that uses a generalized notion of orthogonality for the de Broglie family $e^{\frac{i}{\hbar} p x}$.

\begin{proof}
    Plugging def.~\ref{eq:wigner} into eq.~\eqref{eq:kernel_reconstruction} yields
    \begin{align}
        K_\psi(x,y) &= \frac{1}{2\pi \hbar} \int_{\R}  \left( \int_{\R} e^{-\frac{i}{\hbar} pz} \, \psi\left(\frac{x + y + z}{2} \right) \, \overline{\psi\left( \frac{x + y - z}{2} \right)} \, \mathrm dz \right) \mathrm dp
    \end{align}
    where the inner integral over $z$ can be interpreted as the application of the tempered distributions $\eta_p = \langle e^{\frac{i}{\hbar} pz} | \, \cdot \, \rangle$ parametrized by $p$ to the function $a(z) = \psi\left(\frac{x + y + z}{2} \right) \, \overline{\psi\left( \frac{x + y - z}{2} \right)}$. The result is a function of $p$ evaluated by the tempered distribution $\eta_{x-y}$. In formulas,
    \begin{align}
        K_\psi(x,y) = \frac{1}{2\pi \hbar} \, \eta_{x-y} \big[ \underbrace{p \to \eta_p^\ast}_{\eta^\ast} [a] \big] \; .
    \end{align}
    By delaying the evaluation at $a$ we can generalize the inner product 
    \begin{align}
        K_\psi(x,y) = \frac{1}{2\pi \hbar} \, \langle \eta | \eta_{x-y} \rangle [a] = \delta_{x-y}[a] = a(x-y) \label{eq:orthogonal}
    \end{align}
    and use the fact that $\eta$ is an orthogonal family in that generalized sense (see Refs.~\cite{carfi01, carfi15}) to simplify this to
    \begin{equation}
        K_\psi(x,y) = \psi\left(\frac{x + y + x - y}{2} \right) \, \overline{\psi\left( \frac{x + y - (x - y)}{2} \right)} = \psi(x) \, \overline{\psi(y)}
    \end{equation}
    as expected.
\end{proof}

\subsection{The conjugation identity}
We justify the identity
\begin{equation}\label{eq:conj_identity}
	U_\alpha^\dagger e^{\frac{i}{\hbar} s X} U_\alpha
	=
	e^{\frac{i}{\hbar} s \, U_\alpha^\dagger X U_\alpha}.
\end{equation}

First, conjugation respects products:
\begin{equation}
    U_\alpha^\dagger X^n U_\alpha = (U_\alpha^\dagger X U_\alpha)^n
    \qquad \forall n \in \N_0.
\end{equation}
Second, the exponential of an operator is defined by the power series
\begin{equation}
    e^X = \sum_{n=0}^\infty \frac{X^n}{n!} \; .
\end{equation}
Applying conjugation term by term, we get
\begin{equation}
    U_\alpha^\dagger e^{\frac{i}{\hbar} s X} U_\alpha
    = \sum_{n=0}^\infty \frac{1}{n!} U_\alpha^\dagger \left(\frac{i}{\hbar} s X \right)^n U_\alpha
    = \sum_{n=0}^\infty \frac{1}{n!} \left( \frac{i}{\hbar} s \, U_\alpha^\dagger X U_\alpha \right)^n \; ,
\end{equation}
which is exactly
\begin{equation}
    e^{\frac{i}{\hbar} s \, U_\alpha^\dagger X U_\alpha} \; .
\end{equation}

This argument is fully legitimate on $\mathcal S(\mathbb R)$, since
$\mathcal S(\mathbb R)$ is stable under $X$, $H$, $U_\alpha$, and all powers appearing in the series.

\subsubsection{Alternative proof}
Alternatively, use the fact that $X$ has an eigenbasis of Dirac delta $\delta_q$ when extended to $\mathcal{S}'$.
\begin{equation}
	X \delta_q = q \delta_q
\end{equation}
and can be rotated by the unitary evolution into an eigenbasis $\beta_{\alpha, q} = U_\alpha^\dagger \delta_q$ for (the extension of) the conjugated operator $X(\alpha) = U_\alpha^\dagger X U_\alpha$.
\begin{equation}
	X(\alpha) \beta_{\alpha, q} = U_\alpha^\dagger X U_\alpha U_\alpha^\dagger \delta_q = U_\alpha^\dagger X \delta_q = U_\alpha^\dagger q \delta_q = q \beta_{\alpha, q}
\end{equation}
It is now easy to see that the left- and right-hand side of eq.~\eqref{eq:conj_identity} act in the same way on this basis (and therefore everywhere).

The left-hand side evaluates to
\begin{align}
		U_\alpha^\dagger e^{\frac{i}{\hbar} s X} U_\alpha \beta_{\alpha, q} &= U_\alpha^\dagger e^{\frac{i}{\hbar} s X} U_\alpha U_\alpha^\dagger \delta_q \\
		&= U_\alpha^\dagger e^{\frac{i}{\hbar} s X} \delta_q \\
		&= U_\alpha^\dagger e^{\frac{i}{\hbar} s q} \delta_q \\
		&= e^{\frac{i}{\hbar} s q} U_\alpha^\dagger \delta_q \\
		&= e^{\frac{i}{\hbar} s q} \beta_{\alpha, q}
\end{align}

The right-hand side trivially simplifies to
\begin{equation}
	e^{\frac{i}{\hbar} s \, U_\alpha^\dagger X U_\alpha} \beta_{\alpha, q} = e^{\frac{i}{\hbar} s \, X(\alpha)} \beta_{\alpha, q} = e^{\frac{i}{\hbar} s \, q} \beta_{\alpha, q}
\end{equation}
since $\beta_{\alpha, q}$ is an eigenvector of $X(\alpha)$ with eigenvalue $q$.

\subsection{Solution of the Heisenberg equation for the harmonic oscillator}\label{app:harmonicoscillatorheisenberg}
\begin{align}
	A = \frac{1}{2} \left( X^2 + P^2 \right) \, ,\quad
	U_\alpha = e^{-\frac{i}{\hbar} \alpha A} \\
	\mathcal{X}(\alpha) = U_\alpha X U_\alpha^\dagger = \cos \alpha \, X + \sin \alpha \, P \, . \label{eq:freesolution}
\end{align}

The first derivative is
\begin{align}
		\frac{\mathrm dX}{\mathrm d\alpha} &= \frac{\mathrm d}{\mathrm d\alpha} U_\alpha X U_\alpha^\dagger = \frac{\mathrm d}{\mathrm d\alpha} e^{-\frac{i}{\hbar} \alpha A} X e^{\frac{i}{\hbar} \alpha A} \\
		&= -\frac{i}{\hbar} A \left( e^{-\frac{i}{\hbar} \alpha A} X e^{\frac{i}{\hbar} \alpha A} \right) + \left( e^{-\frac{i}{\hbar} \alpha A} X e^{\frac{i}{\hbar} \alpha A} \right) \frac{i}{\hbar} A \\
		&= \left[ X(\alpha) , \frac{i}{\hbar} A \right] = \frac{i}{\hbar} \left[ X(\alpha) , A \right].
\end{align}

The second derivative is
\begin{align}
		\frac{\mathrm d^2X}{\mathrm d\alpha^2} &= \frac{\mathrm d}{\mathrm d\alpha} \frac{i}{\hbar} \left[ X(\alpha) , A \right] = \frac{i}{\hbar} \left[ \frac{\mathrm d}{\mathrm d\alpha} X(\alpha) , A \right] \\
		&= \left[ \left[ X(\alpha) ,\frac{i}{\hbar}  A \right] , \frac{i}{\hbar} A \right] \\
		&= -\frac{1}{\hbar^2} \left[ \left[ X(\alpha) , A \right] , A \right].
\end{align}

The relevant commutators are
\begin{align}
	\left[ X , P \right] &= i \hbar \\
	\begin{split}
		\left[ X, P^2 \right] &= P \left[ X , P \right] + \left[ X , P \right] P \\
		&= P i \hbar + i \hbar P = 2i \hbar P
	\end{split} \\
	\begin{split}
		\left[ X, A \right] &= \left[ X, \frac{1}{2} \left( X^2 + P^2 \right) \right] = \frac{1}{2} \left[ X, X^2 + P^2 \right] \\
		&= \frac{1}{2} \left[ X, P^2 \right] = i \hbar P
	\end{split} \\
	\begin{split}
		\left[ P, X^2 \right] &= X \left[ P , X \right] + \left[ P , X \right] X \\
		&= -X i \hbar - i \hbar X = -2i \hbar X
	\end{split} \\
	\begin{split}
		\left[ P, A \right] &= \left[ P, \frac{1}{2} \left( X^2 + P^2 \right) \right] = \frac{1}{2} \left[ P, X^2 + P^2 \right] \\
		&= \frac{1}{2} \left[ P, X^2 \right] = -i \hbar X
	\end{split} \\
	\left[ U_\alpha , A \right] &= 0 = \left[ U_\alpha^\dagger , A \right] \\
	\begin{split}
		\left[ U_\alpha O U_\alpha^\dagger , A \right] &= U_\alpha O U_\alpha^\dagger A - A U_\alpha O U_\alpha^\dagger = U_\alpha O A U_\alpha^\dagger - U_\alpha A O U_\alpha^\dagger \\
		&= U_\alpha \left[ O , A \right] U_\alpha^\dagger
	\end{split} \\
	\implies \qquad \left[ X(\alpha) , A \right] &= U_\alpha \left[ X , A \right] U_\alpha^\dagger = i \hbar U_\alpha P U_\alpha^\dagger := i \hbar P(\alpha) \\
	\left[ P(\alpha) , A \right] &= U_\alpha \left[ P , A \right] U_\alpha^\dagger = -i \hbar U_\alpha X U_\alpha^\dagger = -i \hbar X(\alpha).
\end{align}

The operator ODE is therefore
\begin{align}
	\frac{\mathrm dX}{\mathrm d\alpha} &= \frac{i}{\hbar} \left[ X(\alpha) , A \right] = P(\alpha) \\
	\begin{split}
		\frac{\mathrm d^2X}{\mathrm d\alpha^2} = \frac{\mathrm dP}{\mathrm d\alpha} &= -\frac{1}{\hbar^2} \left[ i \hbar P(\alpha) , A \right] = -\frac{i}{\hbar} \left[ P(\alpha) , A \right] \\
		&= -X(\alpha).
	\end{split}
\end{align}

Let us propose the following Ansatz: $X(\alpha) = a(\alpha) X + b(\alpha) P$.
\begin{align}
	\frac{\mathrm dX}{\mathrm d\alpha} &= \frac{\mathrm da}{\mathrm d\alpha} X + \frac{\mathrm db}{\mathrm d\alpha} P = P(\alpha) \\
	\frac{\mathrm d^2X}{\mathrm d\alpha^2} &= \frac{\mathrm d^2a}{\mathrm d\alpha^2} X + \frac{\mathrm d^2b}{\mathrm d\alpha^2} P = -X(\alpha) = -a(\alpha) X - b(\alpha) P
\end{align}

We therefore arrive at the following ODE and initial conditions for $a$ and $b$.
\begin{align}
	\frac{\mathrm d^2a}{\mathrm d\alpha^2} &= -a(\alpha) &
	\frac{\mathrm d^2b}{\mathrm d\alpha^2} &= -b(\alpha) \\
	\left. \frac{\mathrm da}{\mathrm d\alpha} \right|_{\alpha=0} &= 0 &
	\left. \frac{\mathrm db}{\mathrm d\alpha} \right|_{\alpha=0} &= 1 \\
	a(0) &= 1 &
	b(0) &= 0
\end{align}

The solutions are well-known:
\begin{align}
	a(\alpha) = \cos(\alpha) \quad b(\alpha) = \sin(\alpha) \\
	\implies X(\alpha) = \cos(\alpha) X + \sin(\alpha) P
\end{align}

\subsection{Solution of the Heisenberg equation for the free Hamiltonian}\label{app:freeheisenberg}
Define the Hamiltonian $H = \frac{1}{2m} P^2$ and $U_t = e^{\frac{1}{i\hbar} t H}$. In the Heisenberg picture, operators evolve in time according to
\begin{equation}
	\frac{\mathrm d X}{\mathrm dt} = \frac{1}{i \hbar} \left[ H , X(t) \right],
\end{equation}
which is solved by $U_t^\dagger X(t=0) U_t$. Here we want to show explicitly that $X(t) = X(t=0) + \frac{1}{m} P$.

Instead of following the same technique as for the harmonic oscillator we use here the following trick: $[f(P), X] = \frac{\partial f}{\partial p}(P) [P, X]$.
\begin{align}
		X(t) &= U_t^\dagger X U_t = U_t^\dagger \left( \left[ X , U_t \right] + U_t X \right) = U_t^\dagger \left[ X , U_t \right] + \underbrace{U_t^\dagger U_t}_{1} X \\
		&= X - U_t^\dagger \left[ U_t , X \right] = X - U_t^\dagger \left[ \exp\left( \frac{1}{i \hbar} \frac{t}{2m} P^2 \right) , X \right] \\
		&= X - U_t^\dagger \frac{1}{i \hbar} \frac{t}{m} P U_t \underbrace{[P, X]}_{-i \hbar} \\
		&= X + U_t^\dagger U_t \frac{t}{m} P = X + \frac{t}{m} P
\end{align}
where we have used the fact that $\exp\left( \frac{1}{i \hbar} \frac{t}{2m} P^2 \right)$ commutes with $P$. \qed

\end{document}